# Cavity-QED analysis of InAs quantum-dot single photon production

W. W. Chow,[1] S. Peana, [2] D. I. Herman, [2] K. Y. Lee, [1] A. Cejan, [1] C. Shang,[3] G. Moody[3] , J. E. Bowers[3] and F. Jahnke[4]

[1] Center for Integrated Nanotechnologies, Sandia National Laboratories, Albuquerque, NM 87185-1086, U.S.A.

[2] Sandia National Laboratories, Albuquerque, NM 87185-1086, U.S.A.

[3] Department of Electrical and Computer Engineering, University of California, Santa Barbara, CA 93106 USA.

[4] Institute for Theoretical Physics, University of Bremen, 28334 Bremen, Germany

## Abstract

Cavity quantum electrodynamics is used to study the extent cavity enhancement affects single photon performance of an InAs quantum dot, in terms of emission rate, purity and indistinguishability at different temperatures. Parametric studies show a tradeoff between single photon production rate and purity, as well as between indistinguishability and purity. The results indicate the importance of quantum carrier-photon correlations that yield substantial corrections to mean-field treatments, especially at high light-matter coupling strength, where cavity enhancement is most effective.

## 1. INTRODUCTION

The generation of single photons is important for photonic quantum technologies, such as secure communication, distributed quantum computing, and quantum-enhanced sensing. [1] [2] Advances are being made in single photon generation, in terms of emission rate, purity and indistinguishability. [3] [4] [5] Approaches range from trapped ions and color centers to nonlinear optical processes, such as spontaneous parametric down-conversion and four-wave mixing. [6] [7] [8]

Another strong candidate for single photon generation is a semiconductor quantum dot (QD) in an optical micro- or nanocavity. [3] [9] [10] High purity emission, indistinguishable photons and entangled photon pairs have been demonstrated in experiments. [11] [12] [13] Advances in epitaxial growth now allow site-controlled QD positioning and integration on both III–V and silicon substrates. [14] Microcavity engineering has improved brightness and collection efficiency. [15] [16] Electrical triggering will enable scalable on-chip architectures. [17]

Nevertheless, there are still opportunities to improve QD single photon generation. A vast parameter space provides a broad selection of device configurations for optimizing and tailoring performance. However, exploring potential configurations is currently hindered by very demanding fabrication challenges. This paper describes the development and application of a theory to provide timely exploration of device designs of interest. The theory is based on a cavity-quantum electrodynamics (cQED) treatment of a QD inside a micro- or nanocavity. [18] Numerical modeling provides screening of promising single photon source configurations, prior to making costly experimental investment.

In this paper, the application of the theory focuses on the extent cavity enhancement influences single photon performance of an InAs quantum dot. Of interest are the emission rate, purity and indistinguishability at different temperatures. The results show tradeoffs when optimizing these quantities to tailor to specific application. Also, they indicate importance of quantum carrier-photon correlations that may lead to noticeable corrections to mean-field treatments of purity and indistinguishability.

The rest of this paper is organized as follows. Section 2 presents the coupled equations of motion that are solved to determine single photon emission rate, intensity correlation $g^{(2)}(0)$, and coherence time. Their derivations are described in Appendix I. Section 3 discusses the tradeoff between single photon emission rate and purity. Temperature dependence is taken into account via changes in the dephasing rate. The relation between dephasing rate and temperature is determined using quantum kinetic theory. A discussion of the calculation is in Appendix II. Section 4 discusses the subject of indistinguishability. A relationship is shown between the number of coincidences in a Hong-Ou-Mandel (HOM) interference measurement and the calculated product of emission rate and coherence time. [19] [20] This section also describes a tradeoff between indistinguishability and purity, as well as a limit on indistinguishability at high temperature. Lastly Section 5 summarizes the results and present maps of single photon emission rate, $g^{(2)}(0)$ and indistinguishability for combinations of light-matter couple strength and photon cavity lifetime at different temperatures.

## 2. Theory

The starting point is the Hamiltonian [21] [18]

$$H = \hbar\nu\left(a^\dagger a + \frac{1}{2}\right) + \varepsilon_e c^\dagger c + \varepsilon_h b^\dagger b - i\hbar g\left(b^\dagger c^\dagger a - a^\dagger c\, b\right), \qquad (1)$$

where $\hbar\nu$ is the photon energy, $\varepsilon_e$ and $\varepsilon_h$ are the QD electron and hole ground-state energies, $a^\dagger$ and $a$ are the photon creation and annihilation operators, $c^\dagger$ and $c$ are the creation and annihilation operators for electrons, $b^\dagger$ and $b$ are the corresponding operators for holes. The light-matter coupling strength (vacuum Rabi frequency) is [22]

$$g = \wp \sqrt{\frac{\nu}{\hbar \epsilon_0 n_B^2 V}}\, W(R_{QD}) \,, \tag{2}$$

where $\wp$ is the dipole matrix element, $V$ is the optical mode volume, $\hbar$ is Planck's constant divided by $2\pi$, $\epsilon_0$ is the vacuum permittivity, $n_B$ is the background refractive index, $W$ is the amplitude of the passive optical mode eigenfunction at $R_{QD}$, the location of the QD within the optical cavity. We ignore electronic structure details such as overlapping of electron and hole envelope functions. The photon operators obey commutation relations, while the carrier operators obey anti-commutation relations.

The derivations for the equations of motion are performed in the Heisenberg picture, with

$$i\hbar \frac{d\mathrm{AB}}{dt} = [AB, H] \,, \tag{3}$$

where AB represents a product of photon and QD creation and annihilation operators and [⬚] denotes commutator. Central to the approach is the factorization scheme used on operator products arising from Equation (3). For illustration, in the case of the product $c^\dagger c\, a^\dagger a$, the factorization scheme writes

$$\langle c^\dagger c\, a^\dagger a\rangle = \langle c^\dagger c\, \rangle\langle\, a^\dagger a\rangle + \delta\langle c^\dagger c\, a^\dagger a\rangle \tag{4}$$

which is the sum of the mean field contribution $\langle c^\dagger c\, \rangle\langle\, a^\dagger a\rangle$ and the correlation $\delta\langle c^\dagger c\, a^\dagger a\rangle$. Further detail, including the treatment of the correlations, is described in Appendix I.

The end result are equations of motion for the electron, hole, and photon populations, $n_e = \langle c^\dagger c\rangle$, $n_h = \langle b^\dagger b\rangle$ and $n_p = \langle a^\dagger a\rangle$, respectively, as well as the polarization $p = \langle a^\dagger c\, b\rangle$:

$$\frac{dn_e}{dt} = -2gRe(p) + P(1 - n_e) - \gamma_{nr} n_e - \gamma_{nl} n_e n_h \,, \tag{5}$$

$$\frac{dn_h}{dt} = -2gRe(p) + P(1-n_h) - \gamma_{nr}n_h - \gamma_{nl}n_e n_h \, , \tag{6}$$

$$\frac{dn_p}{dt} = 2gRe(p) - \gamma_c n_p \, , \tag{7}$$

$$\frac{dp}{dt} = -[\gamma + \gamma_c/2 + i(\omega - \nu)]p + gn_e n_h + g(n_e + n_h - 1)n_p$$

$$+g(\delta\langle c^\dagger c a^\dagger a\rangle + \delta\langle b^\dagger b a^\dagger a\rangle + \delta\langle b^\dagger c^\dagger c b\rangle) \tag{8}$$

In the above equations, we introduce a nonradiative carrier loss $\gamma_{nr}$, cavity photon decay rate $\gamma_c$, dephasing rate $\gamma$, and rate of spontaneous emission into free space and all other optical modes $\gamma_{nl}$. For the excitation, we consider carrier injection directly into QD levels at rate $P$, with $(1-n_\sigma)$ accounting for Pauli blocking. When numerically solving Equations (5) – (8), we find that the populations and polarization are sufficiently accurately determined with only the mean field contributions. Therefore, the correlations $\delta\langle c^\dagger c a^\dagger a\rangle$, $\delta\langle b^\dagger b a^\dagger a\rangle$ and $\delta\langle b^\dagger c^\dagger c b\rangle$ in Equation (8) can be neglected in the parametric studies to reduce computational demand.

For single photon purity, we compute the equal-time intensity correlation,

$$g^{(2)}(0) = \frac{\langle a^\dagger a^\dagger a a\rangle}{\langle a^\dagger a\rangle\langle a^\dagger a\rangle} = 2 + \frac{\delta\langle a^\dagger a^\dagger a a\rangle}{n_p^2} \, , \tag{9}$$

where following the factorizing scheme of Equation (4) gives $\langle a^\dagger a^\dagger a a\rangle = n_p^2 + n_p^2 + \delta\langle a^\dagger a^\dagger a a\rangle$ for the numerator. As explained in Appendix I, the equation of motion for the doublet contribution $\delta\langle a^\dagger a^\dagger a a\rangle$ is

$$\frac{d\,\delta\langle a^\dagger a^\dagger a a\rangle}{dt} = -4\gamma_c\,\delta\langle a^\dagger a^\dagger a a\rangle + 4g\,\mathrm{Re}(\delta\langle b^\dagger c^\dagger a^\dagger a a\rangle), \tag{10}$$

which is simulateously solved with Equations (11) - (13):

$$\frac{d\ \delta\langle b^\dagger c^\dagger a^\dagger aa\rangle}{dt} = [-(\gamma + 3\gamma_c)\ + i(\omega - \nu)]\ \delta\langle b^\dagger c^\dagger a^\dagger aa\rangle - 2gp^2$$

$$+2g(n_e + n_p)\delta\langle b^\dagger b a^\dagger a\rangle + 2g(n_h + n_p)\delta\langle c^\dagger c a^\dagger a\rangle$$

$$+g(n_e + n_h - 1)\delta\langle a^\dagger a^\dagger aa\rangle\ , \tag{11}$$

$$\frac{d\ \delta\langle c^\dagger c a^\dagger a\rangle}{dt} = -(\gamma_{nr} + 2\gamma_c)\ \delta\langle c^\dagger c a^\dagger a\rangle -\ 2g\mathrm{Re}[pn_e]$$

$$-\ 2g\mathrm{Re}[\delta\langle b^\dagger c^\dagger a^\dagger aa\rangle]\ , \tag{12}$$

$$\frac{d\ \delta\langle b^\dagger b a^\dagger a\rangle}{dt} = -(\gamma_{nr} + 2\gamma_c)\ \delta\langle b^\dagger b a^\dagger a\rangle -\ 2gRe[pn_h]$$

$$-\ 2gRe[\ \delta\langle b^\dagger c^\dagger a^\dagger aa\rangle]\ , \tag{13}$$

In the numerical solution, we use the steady state values for electron-hole polarization, as well as for the photon and QD populations.

For the coherence time,

$$\tau_{coh} = 2\int_{-\infty}^{\infty} d\tau\ \left|g^{(1)}(\tau)\right|^2 , \tag{14}$$

where

$$g^{(1)}(\tau) = \frac{\langle a^\dagger a(\tau)\rangle_{ss}}{n_p^{ss}} \tag{15}$$

with *ss* indicating that we again consider only the stationary situation. [23] Again, we work in the Hiesenberg picture to derive the equation of motion for $\langle a^\dagger a(\tau)\rangle$. The derivation gives

$$\frac{d\ G^{(1)}}{d\tau} = -\gamma_c\ G^{(1)}\ +\ 2gP', \tag{16}$$

$$\frac{dP'}{d\tau} = [i\ (\nu - \omega_n) -\ (\gamma + \gamma_c)]\ P'\ + g\ (n_e +\ n_h - 1)_{ss}\ G^{(1)}\ , \tag{17}$$

where we factored out the rapid oscillations in $\tau$ by defining $G^{(1)}(\tau) = \langle a^\dagger a(\tau)\rangle_{ss} e^{i\nu\tau}$ and $P(\tau) = \langle a^\dagger b^\dagger(\tau) c(\tau)\rangle_{ss} e^{i\nu\tau}$, as discussed in Ref. [23]. Similar to $g^{(2)}(0)$, Equations (16) and (17) are numerical solved using the steady-state values for $n_p$, $p$, $n_e$ and $n_h$.

### 3. Single photon emission rate and purity

To apply the theory, we consider an InAs QD embedded in a cavity with top and bottom distributed Bragg reflectors (DBRs). Equations (5) – (8) are numerically solved for the steady state populations and polarization. The results are used to solve Equations (10) – (13) for the correlations necessary to determine $g^{(2)}(0)$ with Equation (9).

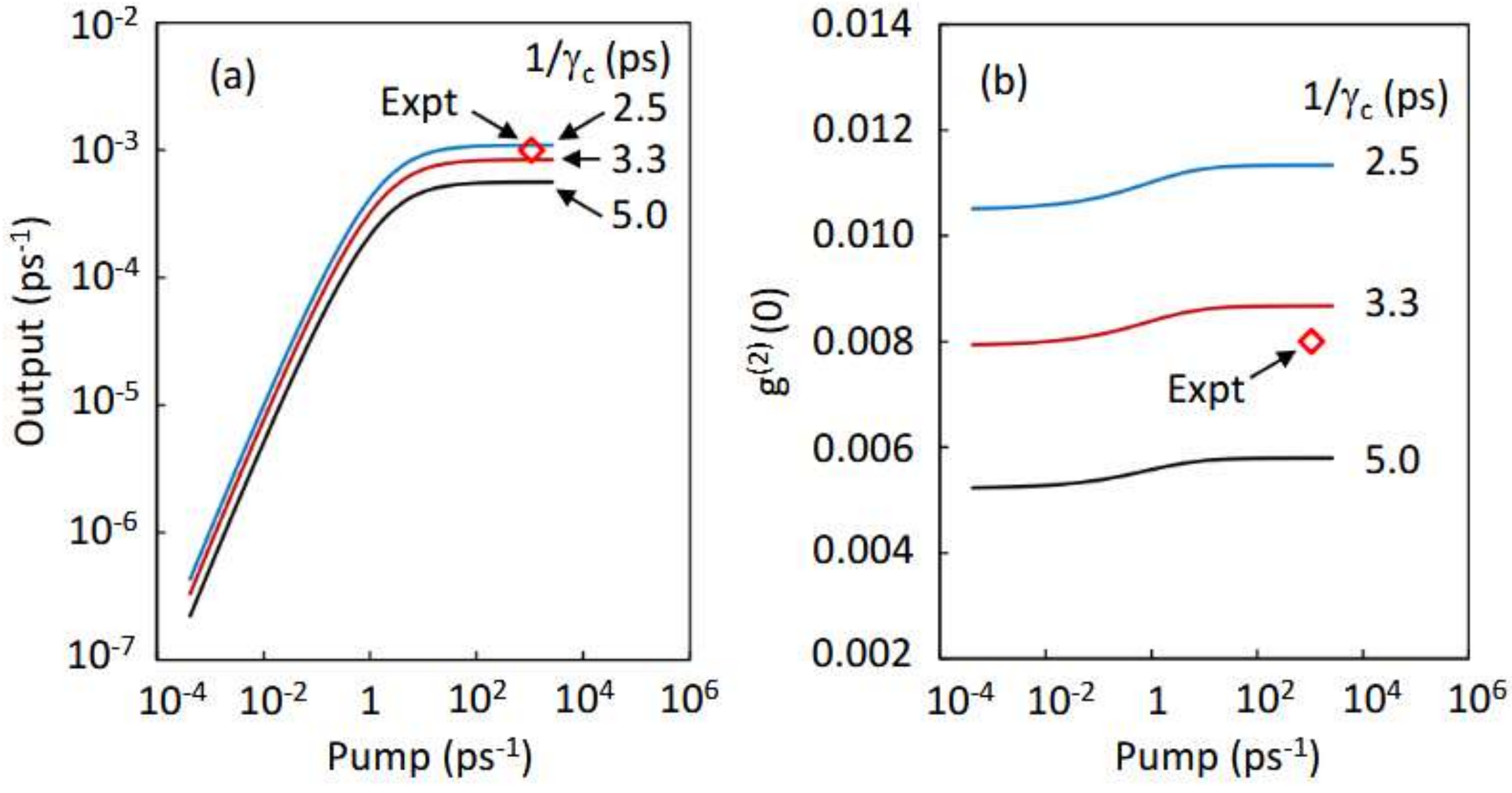

Figure 1. (a) Output and (b) equal-time intensity correlation $g^{(2)}(0)$ versus pump rate. The curves are for cavity photon lifetime $1/\gamma_c$ as indicated. The experimental device gives single photon emission rate of $10^{-3}\ ps^{-1}$ and $g^{(2)}(0) = 0.008$.

Figure 1 shows the computed single photon generation rate $\gamma_c n_p$ and $g^{(2)}(0)$ versus pump $P$ for cavity lifetimes $1/\gamma_c = 5.0, 3.3$ and $2.5\ ps$. To compare to experiment (data points), [24] we assumed a light-matter coupling strength of $g = 0.027\ ps^{-1} = 0.1\ \Omega_{R0}$, where $\Omega_{R0} = 0.27\ ps^{-1}$ is the Rabi frequency computed for a $(\lambda/2n_B)^3$ cavity with QD located at the cavity mode antinode. The other input parameters are wavelength $\lambda = 0.93\ \mu m$, background refractive index $n_B = 3.5$, dipole matrix element $\wp = e \times 0.5\ nm$, $e$ is the electron charge and nonradiative decay rate $\gamma_{nr} = 10^9 s^{-1}$. Also, the spontaneous emission rate into other cavity modes and free space $\gamma_{nl} = 10^{12} s^{-1}$, which gives a spontaneous emission factor $\beta = 10^{-3}$. The experiment was performed at liquid Helium temperature. Based on quantum kinetic calculations for carrier-carrier and carrier-phonon scattering (see Appendix II), we use a dephasing rate $\gamma = 3.9 \times 10^{10}$ for $T = 4\ K$. With increasing excitation, the curves depict increase in emission rate and $g^{(2)}(0)$ until saturation is reached with a completely inverted QD. To further anchor some of the device parameters, we use comparisons of theory and experiments performed for light emitting diodes (LEDs) and lasers, with similarly grown active regions and optical cavities. By increasing QD density, the theory successfully reproduces experimental measurements of excitation dependences of output power, $g^{(2)}(0)$ and coherence time. [25]

To explore the connection between emission rate and single photon purity in more detail, we performed calculations with finer resolution in light-matter coupling strength and cavity lifetime. Changes are made to input parameters used in Figure 1. We consider an active medium that emits at a more useful telecom wavelength of 1.3 μm. In addition, we assume better alignment of the QD and closer in dimension to a $(\lambda/2n_B)^3$ cavity. Figure 2 shows the photon generation rate $\gamma_c n_p$ and $g^{(2)}(0)$ versus cavity photon lifetime $1/\gamma_c$. The curves are for different light-matter coupling strength $g$, which may vary in a device because of misplacement of the QD away from the optical mode antinode. A pump rate of $P = 10^5\,\Omega_{RO}$ injected electron-hole pairs is used to completely invert the QD carrier population, as in single photon experiments. Figure 2 (a) indicates increasing output with increasing cavity lifetime from Purcell enhancement. [26] Also, there is an overall increase in output with increasing light-matter coupling strength $g$. As expected, there is corresponding degradation in purity (increasing $g^{(2)}(0)$ ) with increasing cavity lifetime (Figure 2 (b)). This tradeoff between output and single photon purity has been regarded as a fundamental limitation across all single photon platforms. Although there has been considerable progress in cavity design and fabrication, [11] [17] [27] improving both emission rate and purity simultaneously remains elusive. The curves in Figure 2 (b) suggest hope from a more comprehensive parametric study. In the figure, the black and red curves indicate deviation from monotonic increase in $g^{(2)}(0)$ for $0.3 < 1/\gamma_c < 1\,ps$ and $g \gtrsim 0.8$.

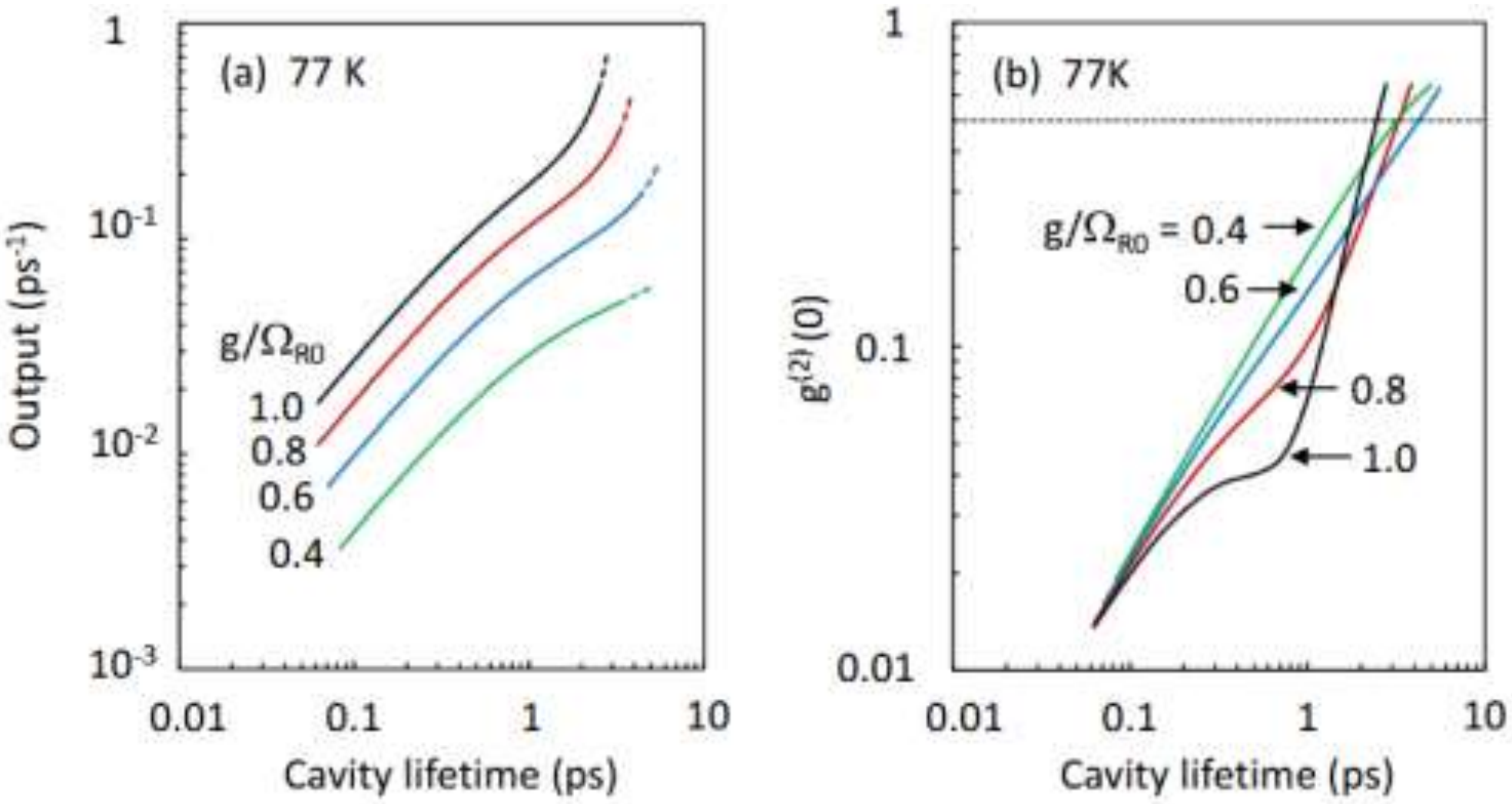

**Figure 2.** Computed (a) output and (b) $g^{(2)}(0)$ versus cavity lifetime for light-matter coupling strength as labeled. The dashed portions in Figure 2 (a) indicate when $g^{(2)}(0) > 0.5$.

The result is interesting and desirable because it indicates the existence of cavity lifetime and light-matter coupling combinations where single photon emission rate increases with reduced purity degradation. We are able to trace the origin to phase correlations contributing to the multiphoton expectation value $\langle a^\dagger a^\dagger aa \rangle$. According to a back-of-the-envelope derivation described in Appendix III, there are certain combinations of $g$ and $1/\gamma_c$ where the dephasing rate of $\delta\langle a^\dagger a^\dagger aa \rangle = \langle a^\dagger a^\dagger aa \rangle - 2n_p^2$ decreases, resulting in reduced randomness in photon emission. Since the effect arises from quantum mechanical correlations, it is not predicted in the often-used estimation of $g^{(2)}(0)$. There, the correlations are neglected through the use of the approximation $\langle a^\dagger a^\dagger aa \rangle = \langle a^\dagger (a\, a^\dagger - 1) a \rangle \approx n_p(n_p - 1)$.

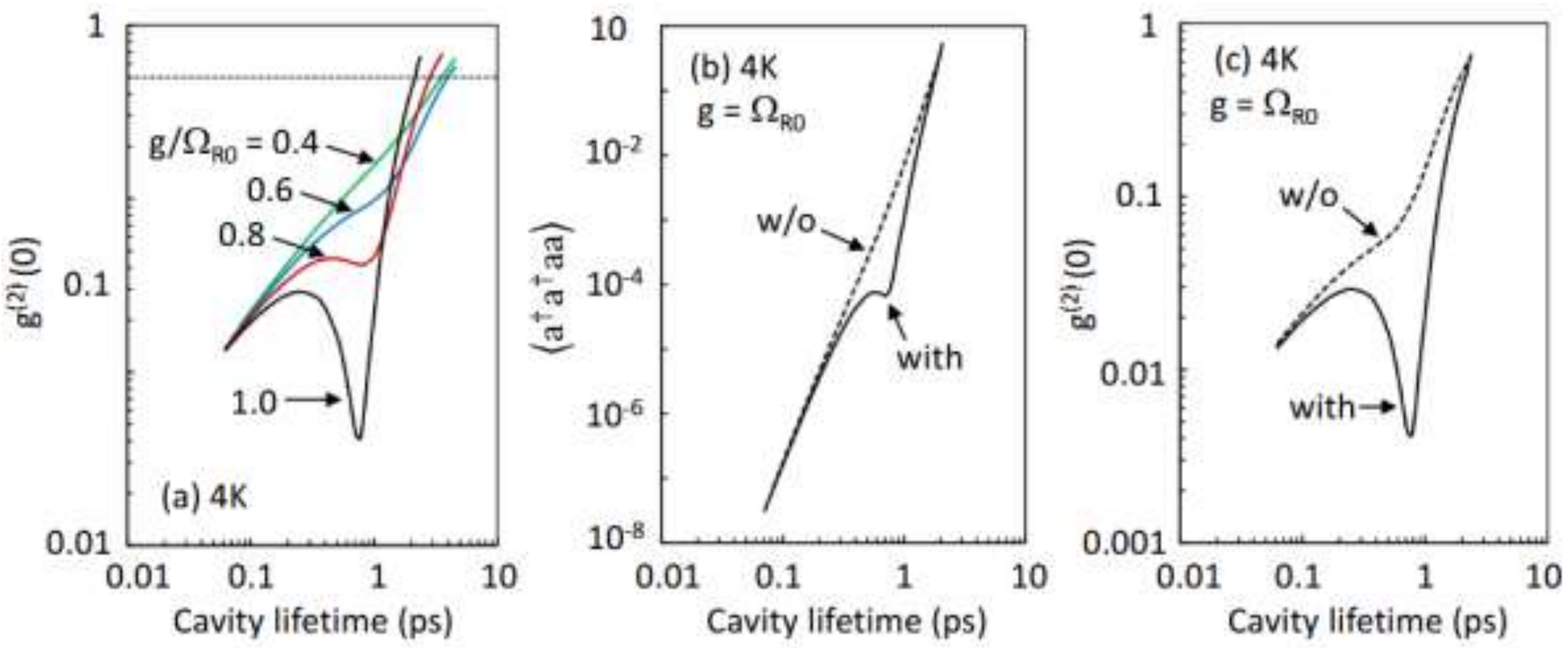

Figure 3. (a) $g^{(2)}(0)$ versus cavity lifetime at liquid He temperature, using $\gamma = 3.9 \times 10^{10}\ s^{-1}$for $T = 4\ K$. (b) Multiphoton expectation value and (c) $g^{(2)}(0)$ versus cavity lifetime computed for $g = \Omega_{R0}$, with and without the $(n_e + n_h - 1)\delta\langle a^\dagger a^\dagger aa \rangle$ contribution in the equation of motion for $\delta\langle b^\dagger c^\dagger a^\dagger aa \rangle$ (solid and dashed curves, respectively).

More definitive confirmation on the role of correlations comes from performing the calculations for lower temperatures. Figure 3 (a) shows accentuated correlation effects resulting in a $g^{(2)}(0)$ dip, i.e. even better purity with higher emission rate. Figure 3 (b) indicates the $g^{(2)}(0)$ dip arises from reduction in the multiphoton expectation value $\langle a^\dagger a^\dagger aa \rangle$. Through process of elimination and the back-of-envelope derivation discussed in Appendix III, we are able to traced the dominant contribution to the $g^{(2)}(0)$ dip to $(n_e + n_h - 1)\delta\langle a^\dagger a^\dagger aa\rangle$, a term in the equation of motion for the polarization-photon-number correlation $\delta\langle b^\dagger c^\dagger a^\dagger aa\rangle$, Equation (11). In Figure 3 (b), the dash curve is computed neglecting this term, which gives an appreciable excitation dependence to the dephasing of $\delta\langle a^\dagger a^\dagger aa\rangle$. Lastly, Figure 3 (c) shows the difference between with and without the contribution in computing $g^{(2)}(0)$.

## 4. Indistinguishability

Indistinguishable single photons are used in quantum information processing. A measure of indistinguishability is the coincident count in a Hong-Ou-Mandel (HOM) interference experiment. [19] Another gauge is the product of photon emission rate and coherence time, or equivalently, the quotient of dephasing time and spontaneous emission time $T_2/(2T_1)$ modified by cavity enhancement. [4] Figure 4 shows the correspondence. On the x-axis is the product $\eta_{ind} \equiv \gamma_c n_p \tau_{coh}$, where the coherence time $\tau_{coh}$ is determined with Equation (14), using the steady-state $g^{(1)}$ from numerically solving Equations (16) and (17). The y-axis is the coherence count $N_c$, also evaluated using $g^{(1)}$ in

$$N_c(\tau_p) = \frac{1}{2}\left[1 - \frac{\int_{-\infty}^{\infty} d\tau\, g^{(1)}(\tau)\, g^{(1)}(\tau - 2\tau_p)}{\int_{-\infty}^{\infty} d\tau\ |g^{(1)}(\tau)|^2}\right], \tag{18}$$

with $\tau_p = 1/(\gamma_c n_p)$. Henceforth, we use $\eta_{ind}$ in discussing indisguishability.

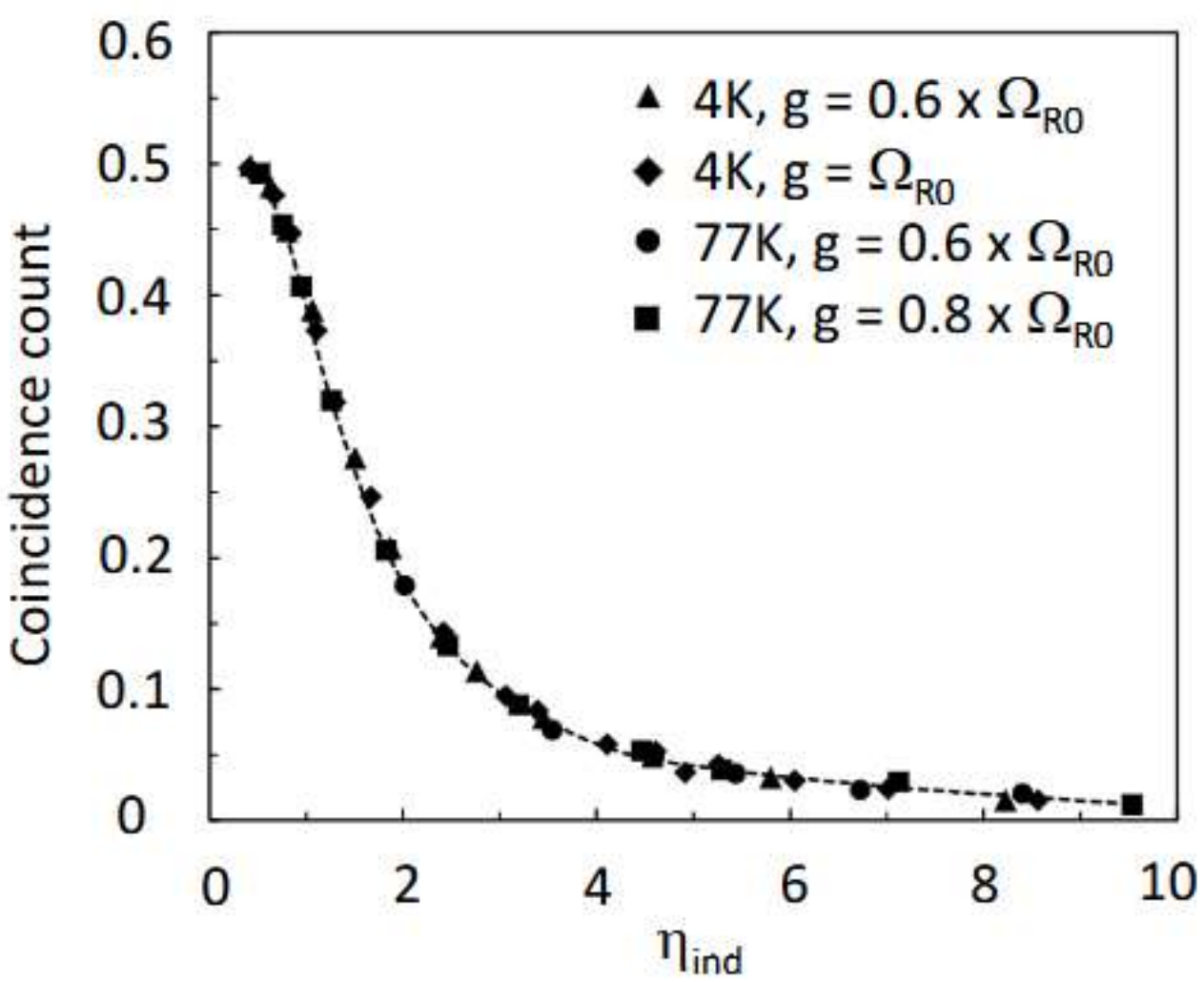

Figure 4. Coincidence count versus product of photon output rate and coherence time. The data points are from calculations at $T = 4\ K$ and $77\ K$.

Figure 5 are plots of coherence time and indisguishability versus cavity lifetime, for temperatures $T = 4\ K$ and $77\ K$. The different curves are for light-matter coupling strengths as indicated. There is a tradeoff between indisguishability and purity, i.e. both $\eta_{ind}$ and $g^{(2)}(0)$ increases with any device configuration. For indistinguishable single photon applications, an experimental configuration has to provide a necessary $\eta_{ind}$ with $g^{(2)}(0)$ not exceeding 0.5. For a curve in Figure 5, the transition between solid to dashed indicates the termination of indistinguishable single photon emission because $g^{(2)}(0) > 0.5$. The differences shown between $T = 4\ K$ and $77\ K$ curves arise from difference in dephasing rates, $\gamma = 3.9 \times 10^{10}\ s^{-1}$ at $T = 4\ K$ to $\gamma = 2.4 \times 10^{11}\ s^{-1}$ at $T = 77\ K$.

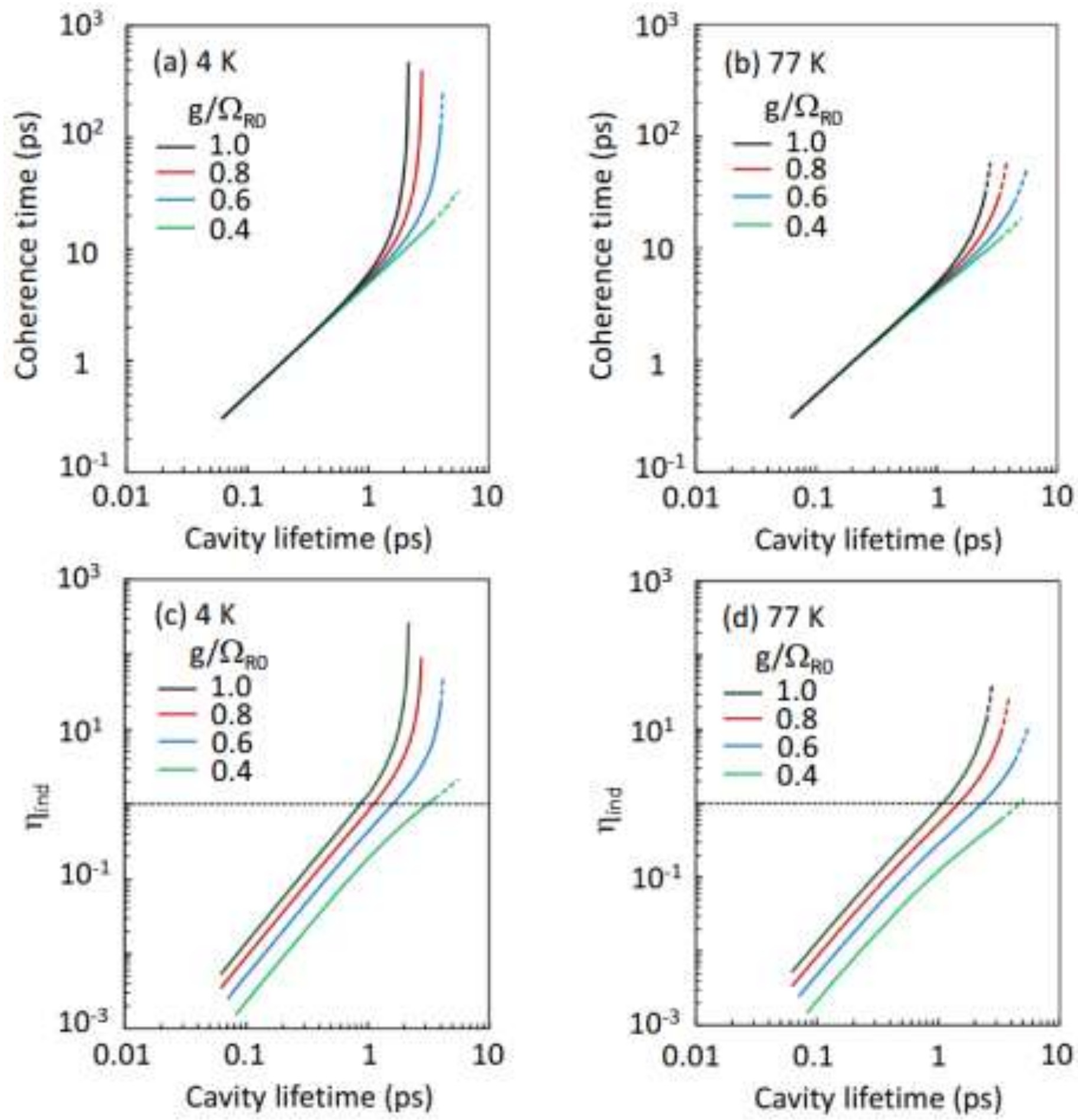

Figure 5. Coherence time and indistinguishability versus cavity lifetime for $T = 4\ K$ and $77\ K$ (left and right columns, respectively). The dashed portion of each curve indicates when $g^{(2)}(0) > 0.5$.

The tradeoff involving purity and indistinguishability constraints the cavity lifetime and light-matter coupling combinations that can provide indistinguishable single photons. Figure 6 are maps of those combinations for $T = 4\ K$ and $77\ K$. At each temperature, the region covered by vertical blue lines give single photon emission (i.e. $g^{(2)}(0) < 0.5$). The horizontal red lines indicate the combinations where $\eta_{ind} > 1$, i.e. where two or more photons are emitted within a coherence time. The region where both sets of lines overlap (cross hatch of red horizontal and blue vertical lines) indicates locations where the combinations of light-matter coupling strength and cavity photon lifetime give indistinguishable single photons. Notable is that for $T = 4\ K$, indistinguishable single photon emission is predicted down to a purity of $g^{(2)}(0) < 0.05$.

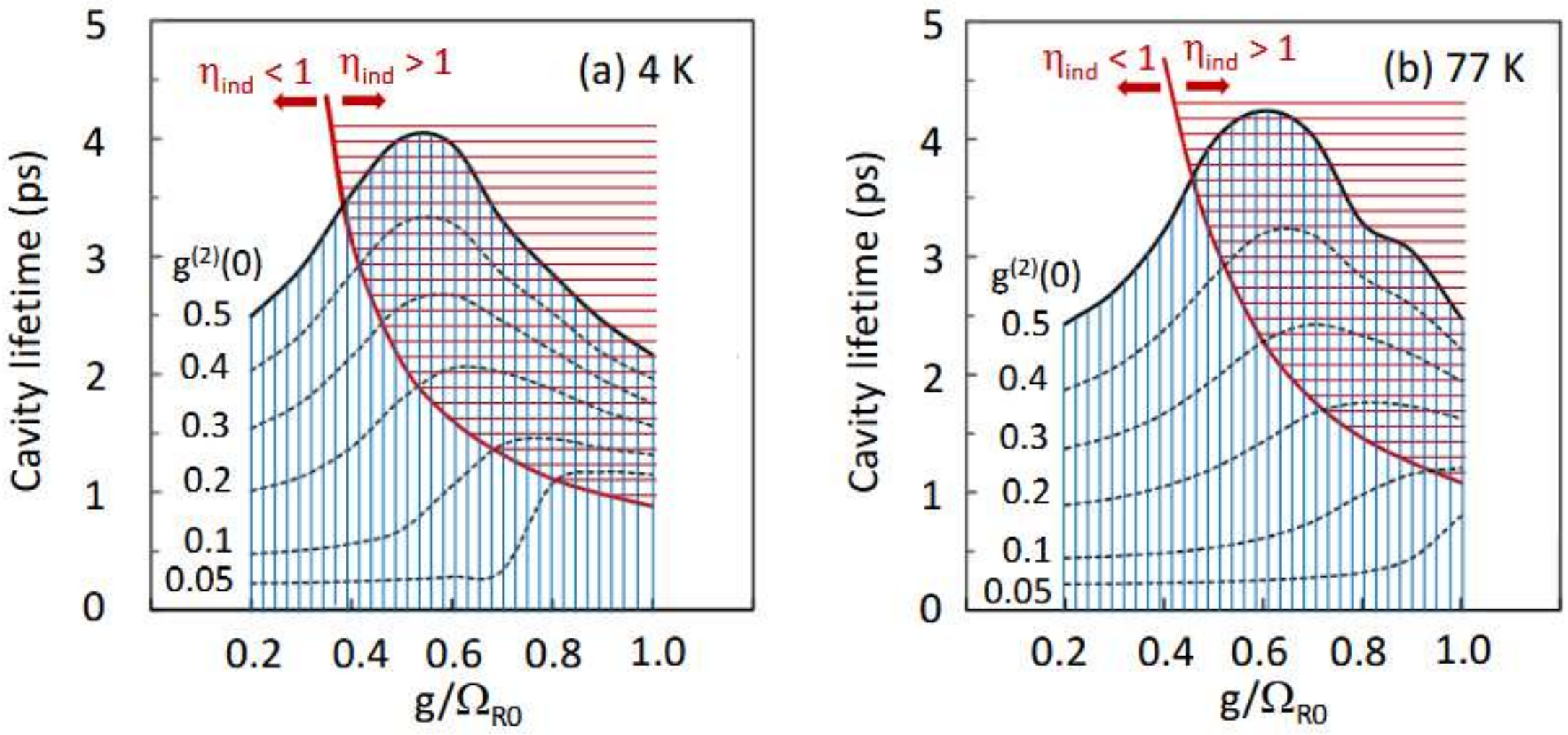

Figure 6. Maps of light-matter coupling strength versus cavity lifetime indicating region of single photon emission (vertical blue lines). Within this broader region is a smaller area (cross hatched red and blue lines) where the conditions give indistinguishable single photons.

Calculations are also performed for temperatures $T = 300\ K$ and $200\ K$, representing room temperature and thermoelectric cooled temperature operation. Figure 7 shows $g^{(2)}(0)$, output and coherence time obtained from solving Equations (5) – (17). The solid and dashed curves are for $T = 300\ K$ and $200\ K$, respectively. At these high temperatures, we use dephasing rate of $\gamma = 1.6\ ps^{-1}$ for $T = 300\ K$ and $1.0\ ps^{-1}$ for $T = 200\ K$. The plots indicate an insensitivity of both $g^{(2)}(0)$ and coherence time to light-matter coupling strength. Comparing Figures 5 and 7 shows appreciable reduction in output and coherence time at high temperature.

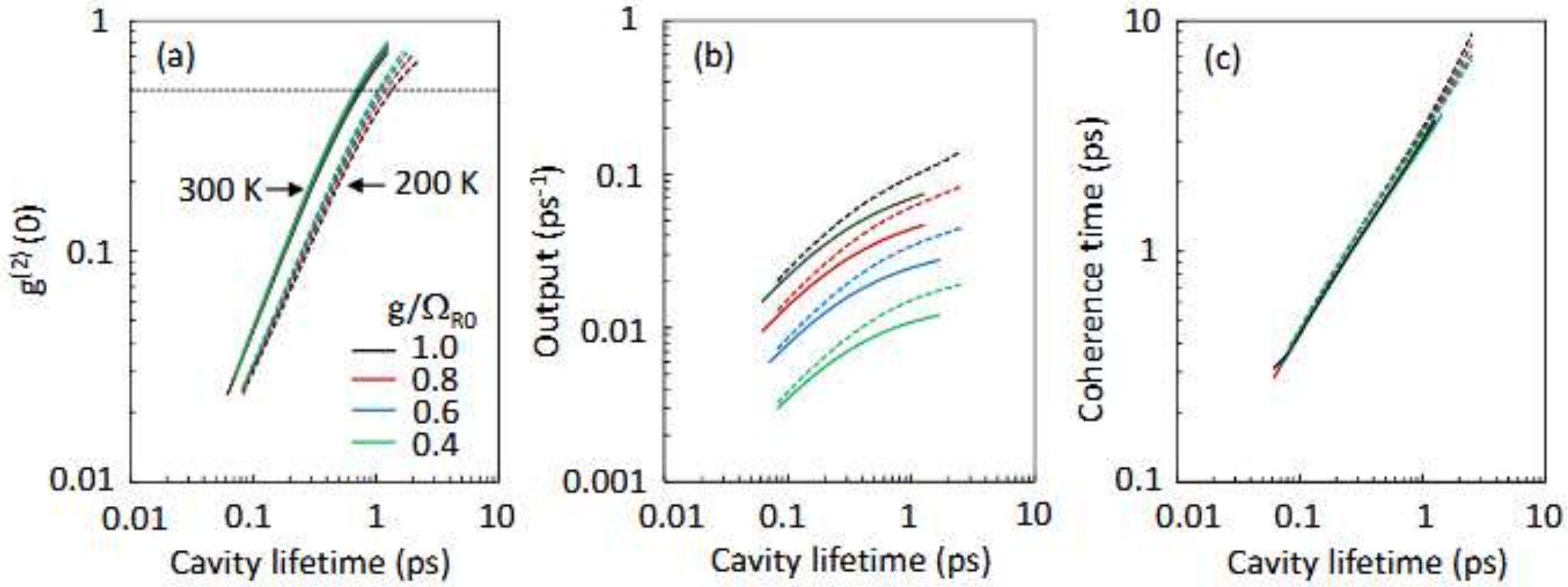

Figure 7. (a) $g^{(2)}(0)$, (b) output and (c) coherence time versus cavity lifetime for $T = 300\ K$ and $200\ K$ (solid and dashed curves). The different colors label the light-matter coupling.

Multiplying the data used in Figures 7 (b) and 7 (c) gives the indistinguishability $\eta_{ind}$ curves in Figure 8. A result of the lower coherence times and emission rates is that $\eta_{ind}$ does not get above unity before $g^{(2)}(0)$ exceeds 0.5, as clearly depicted by the transition from solid to dashed curves.

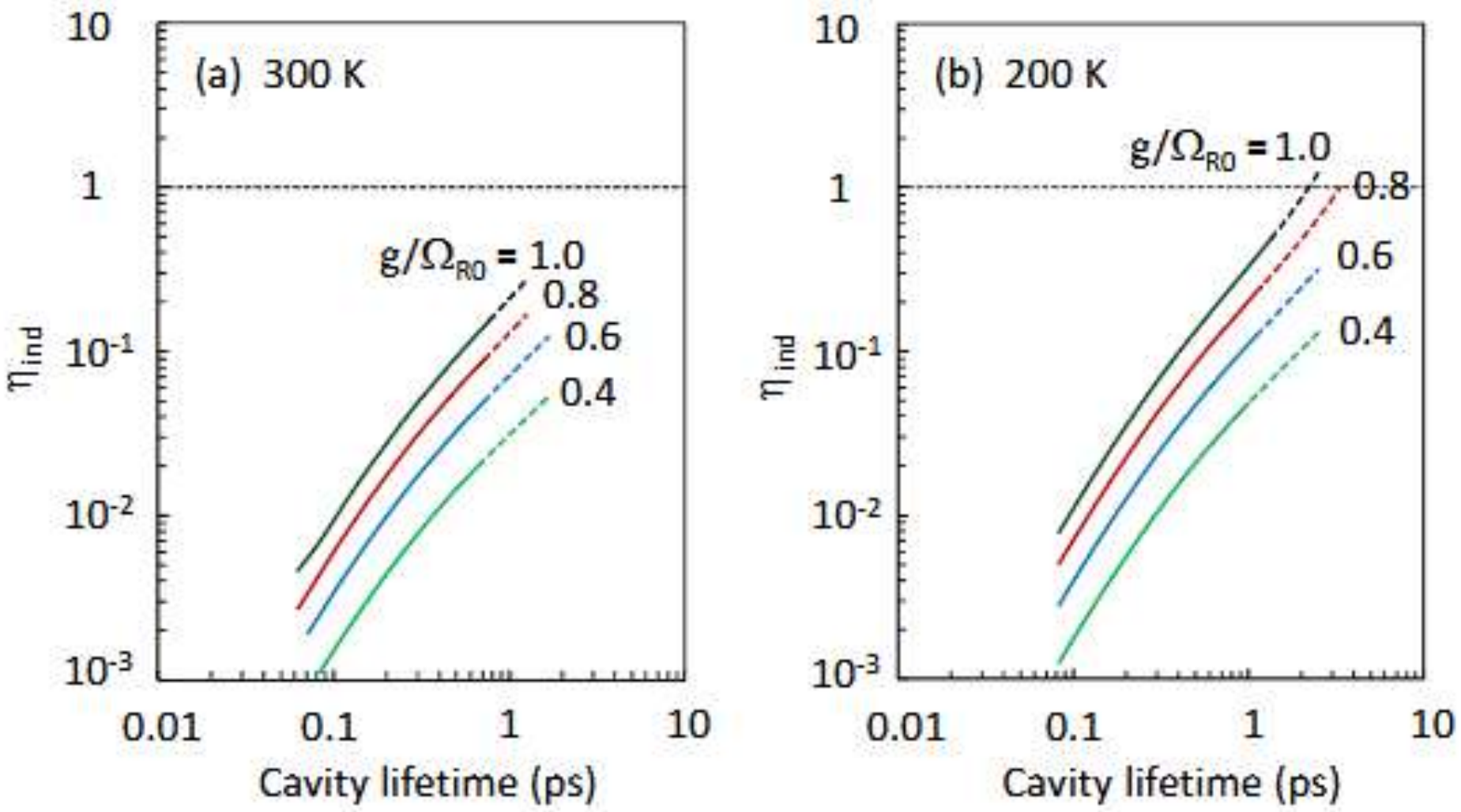

Figure 8. Indistinguishability versus cavity lifetime for (a) $T = 300\ K$ and (b) $200\ K$. The dashed portion of each curve indicates when $g^{(2)}(0) > 0.5$.

## 5. Conclusion

This paper describes a parametric study of single photon generation with a semiconductor quantum dot (QD) in a micro- or nanocavity. The goal is to perform a more comprehensive investigation of the many device configurations than possible with experiment because of extremely challenging and time-consuming fabrication. A cavity-QED theory is developed and applied to provide timely, quantitative evaluation of device configurations for tailoring performance. Numerical modeling allows screening of promising ideas prior to making costly experimental investments.

Our investigation focuses on the extent cavity enhancement affects single photon performance of an InAs quantum dot, in terms of emission rate, purity and indistinguishability. Of importance is dependence on temperature. Figure 9 summarizes the results for four temperatures, with maps of light-matter coupling and cavity lifetime combinations bounded by the $g^{(2)}(0) = 0.5$ contour. The colors, indicating emission rate, show a tradeoff between single photon output and purity. The red curves for $\eta_{ind} = 1$ indicate the subset of light-matter coupling and cavity lifetime combinations giving indistinguishability. At room temperature and thermoelectric cooled temperature, the calculations do not predict indistinguishable single photons.

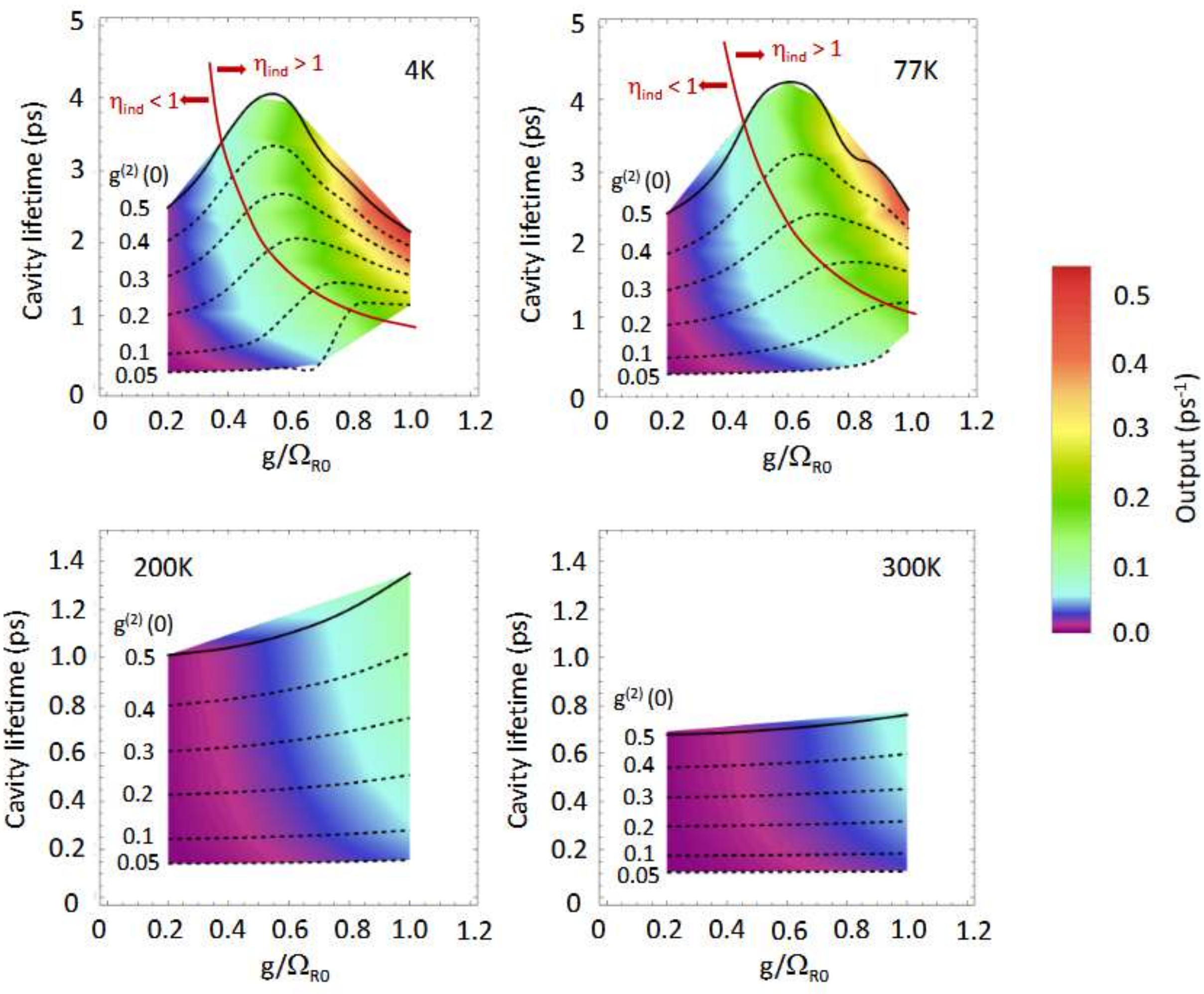

Figure 9. Summary of dependences of single photon emission rate, purity and indistinguishability on combinations of light-matter coupling strength and cavity lifetime. The plots are for temperatures $T = 4\ K$, $77\ K$, $200\ K$ and $300\ K$.

Lastly, our results indicate importance of the quantum mechanical carrier-photon correlations described by the Equations (10) – (17). They lead to deviations in $g^{(2)}(0)$ and $\eta_{ind}$ from mean-field treatments, especially at high light-matter coupling strength, where Purcell enhancement is most effective. The present calculations account for correlations at the doublet level. A future study will examine the influence of high order correlations, which may become important at high light-matter coupling strength. Presently, we treat resonant excitation of the QD. A future study will consider excitation at the embedding quantum well states and time dependences from $\pi$ pulse excitation, especially on changes to indistinguishability.

## ACKNOWLEDGEMENT

This work was performed in part at the Center for Integrated Nanotechnologies, an Office of Science User Facility operated for the U.S. Department of Energy (DOE) Office of Science. Sandia National Laboratories is a multimission laboratory managed and operated by National Technology & Engineering Solutions of Sandia, LLC, a wholly owned subsidiary of Honeywell International, Inc., for the U.S. DOE's National Nuclear Security Administration under Contract No. DE-NA-0003525. The views expressed in the article do not necessarily represent the views of the U.S. DOE or the United States Government. This work was also funded by the NSF Quantum Foundry at UCSB (Grant No. DMR-1906325).

## References

[1] J. W. Silverstone, J. L. Bonneau and M. G. Thompson, "Silicon quantum photonics," *IEEE J Selected Topics in Quantum Electron.,* p. 390, 2016.

[2] S. Slussarenko and G. J. Pryde, "Photonic quantum information processing: A concise review," *Applied Physics Reviews,* vol. 6, p. 041303, 2019.

[3] C. Nawrath, F. Olbrich, M. Pauli, S. L. Portalupi, M. Jetter and P. Michler, "Coherence and indistinguishability of highly pure single photons from non-resonantly and resonantly excited telecom C-band quantum dots," *Appl Phys Lett,* vol. 15, p. 023103, 2019.

[4] L. Husel, J. Scherzer and et al., "Cavity-enhanced photon indistingushablility at room temperature and telecom wavelengths," *Nat. Commun,* vol. 15, p. 3989, 2024.

[5] C. L. Phillips, A. J. Brash, M. Godsland and et al., "Pucell-enhanced single photons at telecom wavelengths from a quantum dot in a photonic crystal cavity," *Sci Rep,* vol. 14, p. 4450, 2024.

[6] W. Redjem, Y. Zhiyenbayev, W. Qarony and et al., "All-silicon quantum light source by embedding an atomic emissive center in a nanophotonic cavity," *Nat Commun,* vol. 14, p. 3321, 2023.

[7] F. Kaneda, K. Garay-Palmett, A. B. U'Ren and et al., "Heralded single-photon source utilizing highly nondegenerate, spectrally factorable spontaneous parametric downconversion," *Opt Express,* vol. 24, p. 10733, 2016.

[8] G. Moody, L. Chang, T. J. Steiner and J. E. Bowers, "Chip-scale nonlinear photonics for quantum light generation," *AVS Quantum Sci,* vol. 2, p. 041702, 2020.

[9] N. Hauser, M. Bayerbach, J. Kaupp and et al., "Deterministic and highly indistinguishable single photons in the telecom C-band," *Nat Commun,* vol. 17, p. 537, 2025.

[10] P. Holewa, D. A. Vajner, E. Zieba-Ostoj and et al., "High-throughput quantum photonic devices emitting indistinguishable photons in the telecom C-band," *Nat Commun,* vol. 15, p. 3358, 2024.

[11] P. Michler, A. Kiraz, C. Becher and et al., "A quantum dot single-photon turnsile device," *Science,* vol. 290\, p. 2282, 2000.

[12] C. Santori, M. Pelton, G. Solomon and et al., "Triggered single photons from a quantum dot," *Phys Rev Lett,* vol. 86, p. 1502, 2001.

[13] Y.-M. He, Y. He, Y.-J. Wei and et al., "On-demand semiconductor single-photon source with near-unity indistinguishability," *Nat Nanotechnol,* vol. 8, p. 213, 2013.

[14] X. Zhao, X. Chen, Y. Bao and et al., "Growth optimization of site controlled single InAs nanoisland arrays on patterned substrates," *Materials Science in Semiconductor Processing,* vol. 208, p. 110524, 2026.

[15] C. Phillips, A. J. Brash, M. Godsland and et al., "Purcell-enhanced single photons at telecom wavelengths from a quantum dot in a photonic crystal cavity," *Scientific Reports,* vol. 14, p. 4450, 2024.

[16] P. Senellart, G. Solomon and A. White, "High-performance semiconductor quantum-dot single-photon sources," *Nat Nanotechnol,* vol. 12, p. 1026, 2017.

[17] I. Aharonovich, D. Englund and M. Toth, "Solid-state single-photon emitters," *Nat Photonics,* vol. 10, p. 631, 2016.

[18] W. W. Chow and S. Reitzenstein, "Quantum-optical influences in optoelectronics - An introduction," *Appl Phys Rev,* vol. 5, p. 041302, 2018.

[19] C. K. Hong, Z. Y. Ou and L. Mandel, "Measurement of subpicosecond time intrvals between two photoncs by interference," *Phys Rev Lett,* vol. 59, p. 2044, 1987.

[20] L. Husel, J. Trapp, J. Scherzer and et al., "Cavity-enhanced photon indistinguishability at room temperature and telecom wavelengths," *Nat Commun,* vol. 15, p. 3989, 2024.

[21] C. Gies, J. Wiersig, J. Lorke and F. Jahnke, "Semiconductor model for quantum-dot based microcavity lasers," *Phys Rev A,* vol. 75, p. 013803, 2007.

[22] W. W. Chow, F. Jahnke and C. Gies, "Emission properties of nanolasers during the transition to lasing," *Light: science and applications,* vol. 3, p. e202, 2014.

[23] S. Ates, C. Gies, S. M. Ulrich and et al., "Influence of the spontaneous optical emission factor on the first-order coherence of semiconductor microcavity laser," *Phys Rev B,* vol. 78, p. 155319, 2008.

[24] C. Shang, M. De Gregorio, Q. Buchinger and et al., "Ultra-low density and high performance InAs quantum dot single photon emitters," *APL Quantum,* vol. 1, p. 036115, 2024.

[25] S. Kreinberg, W. W. Chow, J. Wolters and et al., "Emission from quantum-dot high-beta microcavities: transition from spontaneous emission to lasing and the effects of superradiant emitter coupling," *Light: Science and Applications,* vol. 6, p. e17030, 2017.

[26] E. M. Purcell, "Spontaneous emission probabilities at radio frequencies," *Phys. Rev.,* vol. 69, p. 681, 1946.

[27] N. Somaschi, V. Giesz, L. De Santis and et al., "Near optimal single photon sources in the solid state," *Nat Photon,* vol. 10, p. 340, 2015.

[28] W. W. Chow and S. W. Koch, Semiconductor-Laser Fundamentals: Physics fo the Gain Materials, Berlin: Springer-Verlag, 1999.

[29] H. C. Schneider, W. W. Chow and S. W. Koch, "Excitation-induced dephasing in semiconductor quantum dots," *Phys. Rev. B,* vol. 70, pp. 235308-1-235308-4, 2004.

[30] J. Seebeck, T. R. Nielsen, P. Gartner and F. Jahnke, "Polarons in semiconductor quantum dotsand their role in the quantum kinetics of carrier relaxation," *Phys. Rev. B,* vol. 71, pp. 125327-1-125327-6, 2005.

## Appendix I. Derivation of Equations of Motion

The equations in Section 2 are derived following an approach that starts with the single-particle quantities: intracavity photon number $\langle a^\dagger a \rangle$, carrier populations $\langle c^\dagger c \rangle$ and $\langle b^\dagger b \rangle$ and electron-hole polarization $\langle a^\dagger c\, b \rangle$. To illustrate the derivation, we consider $\langle a^\dagger c\, b \rangle$. In the Hiesenberg picture, it evolves according to

$$i\hbar \frac{d a^\dagger c\, b}{dt} = [a^\dagger c\, b, H] \qquad (A1-1)$$

where the Hamiltonian $H$ is given by Eq. (1). During the derivation, the photon commutation and carrier anticommutation relations are used to arrange terms in normal order, e.g.

$$\begin{aligned}
[a^\dagger cb, b^\dagger c^\dagger a\,] &= a^\dagger cbb^\dagger c^\dagger a - b^\dagger c^\dagger aa^\dagger cb \\
&= (1 - b^\dagger b)(1 - c^\dagger c)a^\dagger a - b^\dagger c^\dagger (1 - a^\dagger a)cb \\
&= (1 - c^\dagger c - b^\dagger b + b^\dagger bc^\dagger c)a^\dagger a - b^\dagger c^\dagger cb - b^\dagger c^\dagger cba^\dagger a \\
&= (1 - c^\dagger c - b^\dagger b)a^\dagger a - b^\dagger c^\dagger cb \ . \qquad (A1-2)
\end{aligned}$$

After taking the expectation value, Equation (A1-1) becomes

$$\frac{d\langle a^\dagger cb \rangle}{dt} = -i(\omega - \nu)\langle a^\dagger cb \rangle + g[\langle (c^\dagger c + b^\dagger b - 1)a^\dagger a \rangle + \langle b^\dagger c^\dagger cb \rangle] \ , \qquad (A1-3)$$

where $\omega = (\varepsilon_e + \varepsilon_h)/\hbar$ is the transition frequency. The next step is to use the factorization scheme:

$$\begin{aligned}
\langle c^\dagger ca^\dagger a\, \rangle &= \langle c^\dagger c \rangle \langle a^\dagger a \rangle + \delta \langle c^\dagger ca^\dagger a \rangle \\
\langle b^\dagger ba^\dagger a\, \rangle &= \langle b^\dagger b \rangle \langle a^\dagger a \rangle + \delta \langle b^\dagger ba^\dagger a \rangle \\
\langle b^\dagger c^\dagger cb \rangle &= \langle b^\dagger b \rangle \langle c^\dagger c \rangle + \delta \langle b^\dagger c^\dagger cb \rangle \ . \qquad (A1-4)
\end{aligned}$$

The result is Equation (8) in the main text,

$$\frac{d\langle a^\dagger cb\rangle}{dt} = -i(\omega-\nu)\langle a^\dagger cb\rangle + g\big(\langle c^\dagger c\rangle + \langle b^\dagger b\rangle - 1\big)\langle a^\dagger a\rangle + g\langle b^\dagger b\rangle\langle c^\dagger c\rangle$$

$$+g(\delta\langle c^\dagger c a^\dagger a\rangle + \delta\langle b^\dagger b a^\dagger a\rangle + \delta\langle b^\dagger c^\dagger cb\rangle) \qquad (8)$$

where the first line are the mean-field contributions and the second line are the quantum correlations involving electron, hole and photon populations. A result of optical field quantization is that the equations of motion for single-particle contains correlations from two particle products (doublets). An equation of motion for the doublets, in turn, has correlations involving three particles (triplet), and so on.

To show how the correlations are treated, we consider $\delta\langle c^\dagger c a^\dagger a\rangle$. Base on the factorization scheme,

$$\langle c^\dagger c a^\dagger a\rangle = \langle c^\dagger c\rangle\langle a^\dagger a\rangle + \delta\langle c^\dagger c a^\dagger a\rangle \quad , \qquad (A1-5)$$

the equation of motion for $\delta\langle c^\dagger c a^\dagger a\rangle$ is

$$\frac{d\delta\langle c^\dagger c a^\dagger a\rangle}{dt} = \frac{d\langle c^\dagger c a^\dagger a\rangle}{dt} - \langle c^\dagger c\rangle\frac{d\langle a^\dagger a\rangle}{dt} - \frac{d\langle c^\dagger c\rangle}{dt}\langle a^\dagger a\rangle \ . \qquad (A1-6)$$

Following steps similar to those in getting Equation (8), we have:

$$\frac{d\langle c^\dagger c a^\dagger a\rangle}{dt} = -g\big(\langle b^\dagger c^\dagger a\rangle\langle a^\dagger a\rangle + \langle a^\dagger cb\rangle\langle a^\dagger a\rangle\big)$$

$$-g\big(\delta\langle b^\dagger c^\dagger a^\dagger aa\rangle + \delta\langle cba^\dagger a^\dagger a\rangle\big) \quad , \qquad (A1-7)$$

$$\frac{d\langle c^\dagger c\rangle}{dt}\langle a^\dagger a\rangle = -g\big(\langle b^\dagger c^\dagger a\rangle + \langle a^\dagger cb\rangle\big)\langle a^\dagger a\rangle \quad , \qquad (A1-8)$$

$$\langle c^\dagger c\rangle\frac{d\langle a^\dagger a\rangle}{dt} = g\langle c^\dagger c\rangle\big(\langle b^\dagger c^\dagger a\rangle + \langle a^\dagger cb\rangle\big) \quad , \qquad (A1-9)$$

Using Equations (A1-7) - (A1-9) in Eq. (A1-6), we get

$$\frac{d\delta\langle c^{\dagger}ca^{\dagger}a\rangle}{dt} = -\langle c^{\dagger}c\rangle g\big(\langle b^{\dagger}c^{\dagger}\mathrm{a}\rangle + \langle a^{\dagger}cb\rangle\big)$$

$$-g\big(\delta\langle b^{\dagger}c^{\dagger}a^{\dagger}aa\rangle + \delta\langle cba^{\dagger}a^{\dagger}a\rangle\big) \qquad (A1-10)$$

This leads to Equation (12), where we assume an effective dephasing rate based on the singlet contributions.

**Appendix II. Dephasing rate**

To include relaxation and dephasing in the equations of motion derived in Appendix I, we add contributions from carrier-carrier and carrier-phonon scattering. For example, the polarization equation of motion becomes,

$$\frac{d\langle a^\dagger cb\rangle}{dt} = -i(\omega-\nu)\langle a^\dagger cb\rangle + g\left(\left\langle c^\dagger c\right\rangle + \left\langle b^\dagger b\right\rangle - 1\right)\left\langle a^\dagger a\right\rangle + g\langle b^\dagger b\rangle\langle c^\dagger c\rangle$$

$$+g(\delta\langle c^\dagger c a^\dagger a\rangle + \delta\langle b^\dagger b a^\dagger a\rangle + \delta\langle b^\dagger c^\dagger cb\rangle)$$

$$+S_0^{c-c} + S_0^{c-p}, \qquad (A2-1)$$

where $S_0^{c-c}$ and $S_0^{c-p}$ come from continuing the Cluster expansion to the 2nd level in Coulomb correlations. [28]

For carrier-carrier scattering, [29]

$$S_0^{c-c} = -p\frac{2}{\hbar}\sum_{\sigma,\rho=e,h}\sum_{k}\sum_{q\neq 0} V_q^2\, g(-\delta\varepsilon_\sigma)\begin{bmatrix}(1-n_\sigma)\left(1-n_{\rho,k}\right)n_{\varrho,k-q}\\ +n_\sigma n_{\rho,k}\left(1-n_{\varrho,k-q}\right)\end{bmatrix}, \qquad (A2-2)$$

where the scattering partners are the momentum $k$-states of the embeding quantum well, $g(x) = \pi\delta(x) + iP(1/x)$ and $\delta\varepsilon_\sigma = \varepsilon_\sigma + \varepsilon_{\rho k} - \varepsilon_{\sigma,k+q} - \varepsilon_{\rho,k-q}$. For carrier-phonon scattering, we use a non-perturbative, quantum-kinetic approach adapted from treating carriers and phonons as composite polarons. [30] The approach gives

$$S_0^{c-p} = -p\left(\Gamma_0^e + \Gamma_0^h\right), \qquad (A2-3)$$

which is evaluated after numerically performing the time integrations:

$$\Gamma_0^\sigma = \int_0^\infty dt' \sum_{\rho=e,h}\sum_{k}\left|\frac{M_{0,k}^{\sigma\rho}}{\hbar}\right|^2 G_k^\rho(t')\, G_0^\sigma(t')$$

$$\times\, e^{i(\varepsilon_\sigma-\varepsilon_{\rho k})t'}\begin{Bmatrix}(1-n_\sigma)\left[n_{LO}e^{i\omega_{LO}t'} + (n_{LO}+1)e^{-i\omega_{LO}t'}\right]\\ n_\sigma\left[(n_{LO}+1)e^{i\omega_{LO}t'} + n_{LO}e^{-i\omega_{LO}t'}\right]\end{Bmatrix} \qquad (A2-4)$$

$$\frac{dG_n^\sigma}{dt} = -\int_0^t dt' \left[(n_{LO}+1)e^{-i\omega_{LO}t'} + n_{LO}e^{i\omega_{LO}t'}\right] G_n^\sigma(t-t')$$

$$\times \sum_{\rho=e,h} \sum_{k'} \left|M_{n,k'}^{\sigma\rho}\right|^2 G_{k'}^\rho(t') e^{i(\varepsilon_{\sigma n}-\varepsilon_{\rho,k'})t'} \qquad (A2-5)$$

In the above equations, $\varepsilon_{\rho k}$ is the electron or hole energy in quantum well $k^{th}$ state, $M_{n,k}^{\sigma\rho}$ is the Frolich matrix element for carrier-LO-phonon interaction, $n_{LO}$ is the phonon populaton and $n$ may be 0 or $k$. Figure A2-1 shows the polarization dephasing rate $\gamma$ versus temperature, calculated using Equations (A2-2) to (A2-5). A carrier density of $N = 2 \times 10^{16}\ m^{-2}$ is used to completely invert the QD ground state.

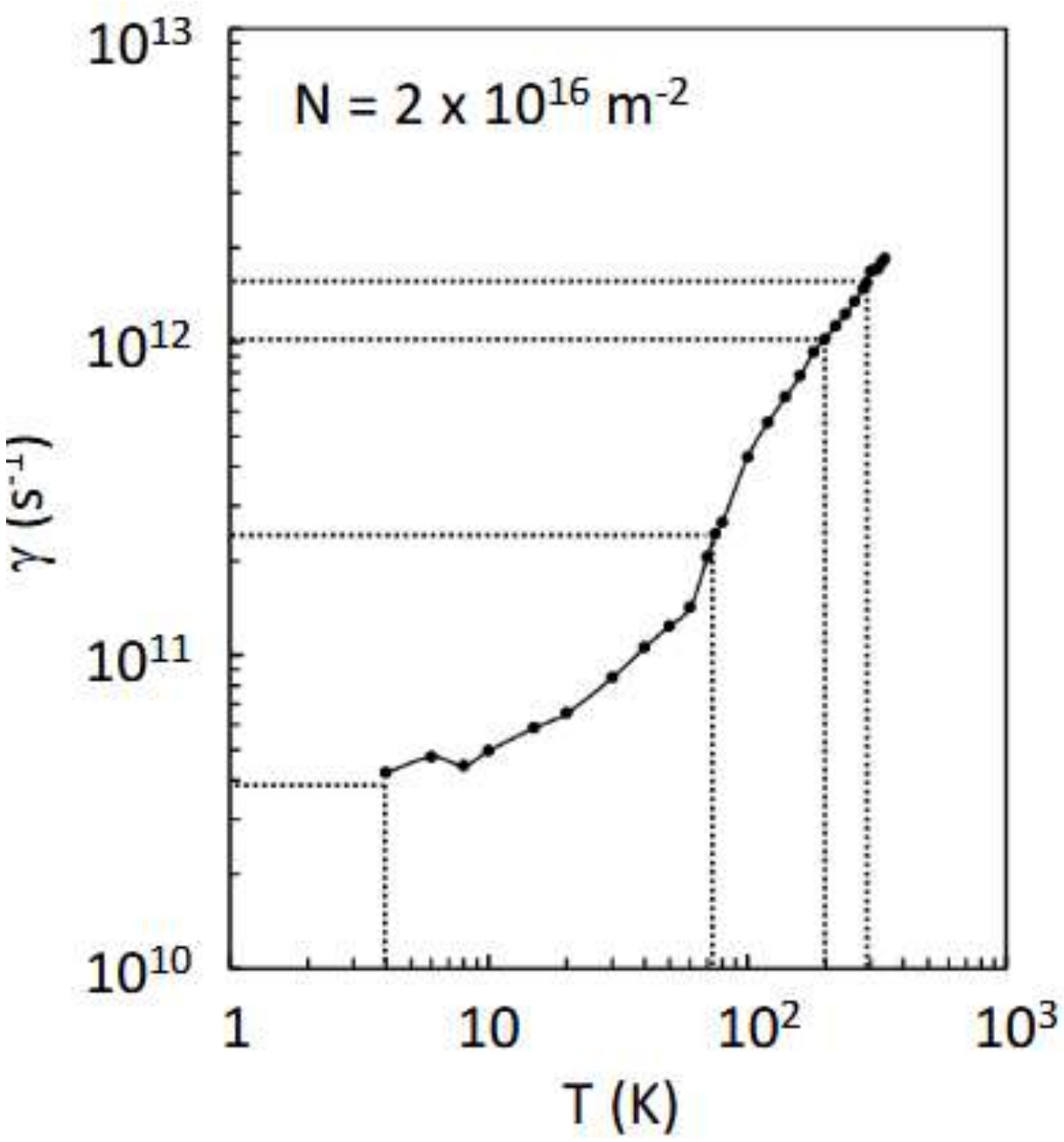

Figure A2-1. Computed polarization dephasing rate ($\gamma$ in Equation 8) versus temperature. For the $T = 4\ K,\ 77K,\ 200\ K$ and $300\ K$, the dash lines indicate $\gamma = 3.9 \times 10^{10} s^{-1}$, $2.4 \times 10^{11} s^{-1}$, $10^{12} s^{-1}$ and $1.6 \times 10^{12} s^{-1}$.

## Appendix III: $g^{(2)}(0)$ dip

Some insight into the non-monotonic increase in $g^{(2)}(0)$ with increasing cavity lifetime may be from the following back-of-envelope derivation. The steady state solutions to Equations (10) and (11) are

$$\delta\langle a^\dagger a^\dagger aa\rangle = \frac{g}{\gamma_c}\mathrm{Re}[\delta\langle b^\dagger c^\dagger a^\dagger aa\rangle]\,, \qquad (A3-1)$$

$$\delta\langle b^\dagger c^\dagger a^\dagger aa\rangle = \frac{g}{(\gamma+3\gamma_c)+i(\omega-\nu)}[(n_e+n_h-1)\delta\langle a^\dagger a^\dagger aa\rangle+\Pi]\,, \qquad (A3-2)$$

where $\hbar\omega = \varepsilon_e + \varepsilon_h$ is the transition energy and $\Pi = (p^2)^* - n_h\delta\langle c^\dagger c a^\dagger a\rangle - n_e\delta\langle b^\dagger b a^\dagger a\rangle$. Substituting Equation (A3-2) into Equation (A3-1) and assuming resonant condition ($\omega = \nu$) for brevity, we get

$$\delta\langle a^\dagger a^\dagger aa\rangle = -2\frac{g}{\gamma_{eff}}\mathrm{Re}[\Pi]\,, \qquad (A3-3)$$

where we introduce an excitation-dependent effective dephasing rate for the multiphoton correlation $\delta\langle a^\dagger a^\dagger aa\rangle$:

$$\gamma_{eff} = \gamma_c - \frac{g^2}{(\gamma+3\gamma_c)}(n_e+n_h-1) \qquad (A3-4)$$

Using Equation (A3-4) in Eq. (9), we have

$$g^{(2)}(0) = 2\left(1 - \frac{g}{\gamma_{eff}}\mathrm{Re}[\Pi]\right) \qquad (A3-5)$$

According to Equation (A3-5), a highly excited, inverted population (hence, higher output) leads to a smaller $\gamma_{eff}$. We note that for the input parameters used in the calculations so far, $\Pi$ to be positive. Hence, a smaller $\gamma_{eff}$ in turn leads to the decrease in $g^{(2)}(0)$ with increasing output. Numerical evaluations of $\Pi$ shows combinations of $g$ and $\gamma_c$ giving greater reduction than others which determine the location of the dip in Figures 2 and 3. Further isolating the contributions is difficult because there is delicate among the terms in $\Pi$. Eliminating one often leads to unphysical results, e.g. to $g^{(2)}(0) < 0$. We will be investigating in greater detail to obtain better understanding.